\documentclass[sn-mathphys]{sn-jnl}

\jyear{2022}%

\usepackage{amsfonts}
\usepackage{amsmath}
\usepackage{booktabs}
\usepackage{multirow}
\usepackage{amssymb}
\usepackage{bbding}
\usepackage{pifont}
\usepackage{wasysym}
\usepackage{color}
\usepackage{siunitx}
\usepackage{subfigure}
\usepackage{graphicx}

\theoremstyle{thmstyletwo}%

\theoremstyle{thmstylethree}%

\newcommand{\bx}{\mathbf{x}}
\newcommand{\by}{\mathbf{y}}

\newcommand{\ba}{\mathbf{a}}
\newcommand{\bbr}{\mathbf{r}}
\newcommand{\mpro}{$M^{\rm pro}$}
\newcommand{\ourDB}{$T_{\rm DB}$}
\newcommand{\ourM}{TamGent}

\begin{document}

\title[Tailoring Molecules for Protein Pockets]{Tailoring Molecules for Protein Pockets: a Transformer-based Generative Solution for Structured-based Drug Design}

\author[1]{\fnm{Kehan} \sur{Wu}}\email{wu\_2018@mail.ustc.edu.cn}
\author[2]{\fnm{Yingce} \sur{Xia}}\email{yingce.xia@microsoft.com}
\author[1]{\fnm{Yang} \sur{Fan}}\email{fyabc@mail.ustc.edu.cn}
\author[2]{\fnm{Pan} \sur{Deng}}\email{pan.deng@microsoft.com}
\author[2]{\fnm{Haiguang} \sur{Liu}}\email{haiguangliu@microsoft.com}
\author[2]{\fnm{Lijun} \sur{Wu}}\email{lijuwu@microsoft.com}
\author[3]{\fnm{Shufang} \sur{Xie}}\email{shufangxie@ruc.edu.cn}
\author[2]{\fnm{Tong} \sur{Wang}}\email{watong@microsoft.com}
\author[2]{\fnm{Tao} \sur{Qin}}\email{taoqin@microsoft.com}
\author[2]{\fnm{Tie-Yan} \sur{Liu}}\email{tyliu@microsoft.com}

\affil[1]{University of Science and Technology of China, \orgaddress{\street{No. 96, Jinzhai Road, Baohe District}, \city{Hefei}, \postcode{230026}, \state{Anhui}, \country{China}}}

\affil*[2]{Microsoft Research, \orgaddress{\street{No. 5, Danling Street, Haidian District}, \city{Beijing}, \postcode{100080}, \country{China}}}

\affil[3]{Renmin University of China, \orgaddress{\street{No.59, Zhongguancun Street, Haidian District}, \city{Beijing}, \postcode{100872}, \country{China}}}


\abstract{Structure-based drug design is drawing growing attentions in computer-aided drug discovery. Compared with the virtual screening approach where a pre-defined library of compounds are computationally screened, {\em de novo} drug design based on the structure of a target protein can provide novel drug candidates. In this paper, we present a generative solution named \ourM{} (\textbf{T}arget-\textbf{a}ware \textbf{m}olecule \textbf{gen}erator with \textbf{T}ransformer) that can directly generate candidate drugs from scratch for a given target, overcoming the limits imposed by existing compound libraries. Following the Transformer framework (a state-of-the-art framework in deep learning), we design a variant of Transformer encoder to process 3D geometric information of targets and pre-train the Transformer decoder on $10$ million compounds from PubChem for candidate drug generation. Systematical evaluation on candidate compounds generated for targets from DrugBank shows that both binding affinity and drugability are largely improved. \ourM{} outperforms previous baselines in terms of both effectiveness and efficiency. The method is further verified by generating candidate compounds for the SARS-CoV-2 main protease and the oncogenic mutant KRAS G12C. The results show that our method not only re-discovers previously verified drug molecules , but also generates novel molecules with better docking scores, expanding the compound pool and potentially leading to the discovery of novel drugs.}

\keywords{Structure-based drug design, generative model, Transformer, SARS-CoV-2 main protease, KRAS G12C mutant}

\maketitle

\section{Introduction}\label{sec:introduction}
Drug discovery is a long and costly process, which often takes 10+ years and billions of dollars for a new FDA-approved drug. Structure-based drug design (SBDD) is one of the core tasks in drug discovery~\citep{ProcessSBDD2003}. Usually, in the process of SBDD, hit compounds selection is the first step after protein target identification and will serve as the source for downstream lead compounds and preclinical compounds (PCC) optimization \cite{heifetz2022artificial}. Nowadays, the physicochemical based methods are widely used for hit compounds selection, such as molecular docking \cite{Bajorath2002,Schneider2010}, free-energy perturbation \cite{doi:10.1021/ja512751q}, and molecular dynamics simulation \cite{doi:10.1021/acs.jmedchem.5b01684,Liu2018-ug}. With these methods, the binding scores between targets and each compound in a library are assessed, and the compounds with less promising scores are filtered out. 

However, there are two limitations of physicochemical based methods. First, for large libraries of compounds, it is extremely costly to enumerate all candidates. Although chemical or pharmaceutical knowledge can be applied to focus on a subset of compounds for virtual screening, the procedure and outcome can be very subjective and depending on experts. The second limitation, which is more critical, is that the novelty of the drugs obtained from compound library is limited to chemical knowledge spanned by the compound libraries. In previous studies, the complete chemical space is estimated to  contain over $10^{60}$ compounds \citep{doi:10.1021/ar500432k}, which is beyond the coverage of any existing libraries (e.g., PubChem has about $10^{8}$ compounds, and ZINC contains about $10^{9}$ compounds, etc).

Recently, generative models originated from Artificial Intelligence (AI) research have been applied to SBDD. For a given target, generative models propose candidate compounds, some of which are not covered in existing libraries. For example, Ragoza et al. \cite{liGAN2020} propose a two-stage method for molecule generation, which first generates an intermediate ligand shape that is complementary to the target, and then decodes the shape into a SMILES string. SBMolGen \cite{StructureBasedMolGen2021} uses an RNN model to generate molecular SMILES guided by Monte Carlo tree search, and the corresponding docking score between the target and the generated drugs are used as the feedback signals to re-weight the search process. Luo et al.  \cite{luo2021sbdd} leverage a graph neural network to encode a target and use Markov chain Monte Carlo algorithm to directly sample atoms of a candidate drug in the 3D spatial space. We denote this method as ``3DGen'' in the following sections for clarity.
SECSE \cite{SystemicEvolutionary2021generating} is a hybrid method, which uses both evolutionary algorithms and deep learning to sequentially add fragments up to form complete candidate drugs.

Despite the progress in AI-based SBDD methods, existing methods are still facing several challenges. First, the number of target-drug (alternatively, target-ligand) pairs is limited for model training. For example, Luo et al. \cite{luo2021sbdd} built a dataset with $100k$ target-drug pairs by filtering the CrossDocked dataset \cite{CrossDocked}. The Protein Data Bank (PDB) harbors $186k$ structure entries with $300k$ pocket-ligand pairs. 
However, the number of unique compounds that bind to targets with resolved structures is only 29 thousands (exclude inorganic molecules and molecules with $\le8$ heavy atoms), much smaller compared to the chemical space (i.e., $10^{60}$). With thousands of unique ligands, it is hard to train a compound generator with good generalization ability. Secondly, the 3D information of the amino acids in the target pocket is not fully exploited in most SBDD AI models. Previous work uses the eigenvalues/eigenvectors \cite{EGCM}, convolutional neural networks \cite{liGAN2020} or graph neural networks \cite{luo2021sbdd} for structure encoding. These approaches cannot jointly model all amino acids in a binding pocket. Therefore, we propose \ourM{} (\textbf{T}arget-\textbf{a}ware \textbf{m}olecule \textbf{gen}erator with \textbf{T}ransformer) to tackle the two substantial challenges. For the first challenge, we pre-train a generative Transformer \cite{Transformer} model (the state-of-the-art model in deep learning) on $10M$ compounds from PubChem \cite{Kim2021-th} to learn the universal characteristics of molecules and thus gains stronger generative power. For the second challenge, we propose a variant of Transformer encoder, that can encode the 3D structure information jointly with the amino acid sequence. We stress two major differences compared to standard Transformers: (1) the atomic coordinates of target proteins are encoded by a feed-forward network; (2) the self-attention layer in standard Transformer only uses the similarity of the input representations to adaptively aggregate the information, while we take the positions of amino acids into consideration.
Considering that multiple drugs can bind to the same target, we use the variational auto-encoder (VAE) framework \cite{Bayes-VAE}, a classical technique to handle to one-to-many mapping, in our method.  

We validate \ourM{} from the following aspects. First, we evaluate the binding affinity (calculated by AutoDock Smina \cite{Koes2013-ta}) and drug-likeness of generated candidate compounds targeting given proteins. For protein targets from DrugBank, our method achieves significantly better results than two recently reported methods (3DGen \cite{luo2021sbdd} and SECSE \cite{SystemicEvolutionary2021generating}). Besides successfully generating the drugs as in the DrugBank dataset, our method also generates new compounds with better docking scores than the reference drugs. We then apply the method to generate candidate compounds for two important targets: the main protease (\mpro{}) of SARS-CoV-2 and the KRAS G12C mutant. The results further support the potential of \ourM{} in practical drug design research.

\section{Results}
\subsection{Method Summary} \label{sec:method-summary}
We briefly describe our method in this section and more details are elaborated in the ``Method'' section. 

{\em Preliminaries}: Let $\ba=(a_1,a_2,\cdots,a_N)$ and $\bbr=(r_1,r_2,\cdots,r_N)$ denote the amino acids and their 3D coordinates of a binding pocket respectively, where $N$ is the sequence length and $r_i\in\mathbb{R}^3$ is the centroid of amino acid $i$ ($i$ is an index to label the amino acids around the binding site). 
$a_i$ is a one-hot vector like $(\cdots, 0,0,1,0,\cdots)$, where the vector length is 20 (the number of possible amino acid types) and the only $1$ locates at the position corresponding to the amino acid type.
A binding pocket is denoted as $\bx=(\ba,\bbr)$ and $[N]=\{1,2,\cdots,N\}$. Let $\by=(y_1,y_2,\cdots,y_M)$ denote the SMILES string of the corresponding ligand/drug with length $M$. Our goal is to learn a mapping from $\bx=(\ba,\bbr)$ to $\by$.

\begin{figure}[!htbp]
\centering
\includegraphics[width=0.9\linewidth]{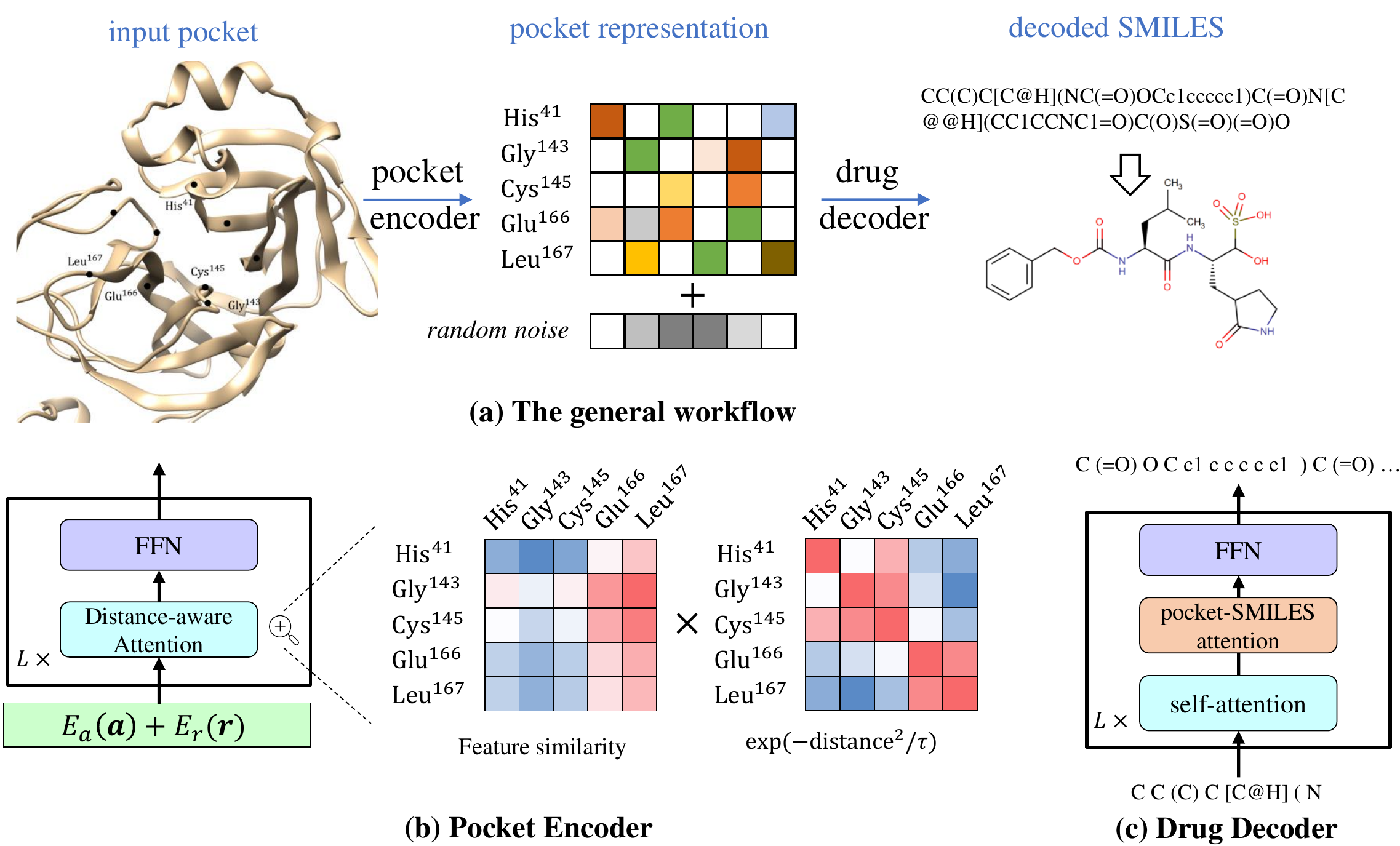}
\caption{\textbf{The architectures of \ourM{}.} (a) The general workflow of our model. A pocket encoder is used to get the high-dimensional representations of amino acids in the pocket. The representations added with random noise are then fed into a drug decoder to generate the SMILES representation of compounds. By adding different noise sets, we obtain different generative results. The drug decoder is pre-trained on $10M$ molecular SMILES. (b) The details of the pocket encoder. The amino acids and the coordinates are mapped into high-dimensional vectors (denoted as $E_a(a)$ and $E_r(r)$ respectively), and iteratively refined by the attention module and the feed-forward layer. The attention adaptively aggregates the amino acid information by considering the production of representation similarity and geometric distances. (c) The architecture of the drug decoder, which outputs the compound SMILES. It is the same as the standard Transformer \cite{Transformer}, which has a self-attention layer to model the preceding SMILES, the pocket-SMILES attention to process pocket representations and the feed-forward layer for non-linear transformation. It is pre-trained on $10$M SMILES from PubChem.
}
\label{fig:networl_arch}
\end{figure}

{\em Framework}: The framework of \ourM{} is shown in Figure \ref{fig:networl_arch}, which is inspired from the variational auto-encoder (VAE) framework~\citep{Bayes-VAE, CVAE, Generative-VAE}. It consists of a pocket encoder, a drug decoder and a VAE encoder (only used in the training phase, not shown in Figure \ref{fig:networl_arch}), all based on the Transformer backbone \cite{Transformer}. A Transformer module consists of a self-attention layer, an encoder-decoder attention layer that only exists in the decoder (i.e., the pocket-SMILES layer in Figure \ref{fig:networl_arch}(c)), and a feed-forward layer (denoted as \texttt{FFN}). Each attention layer outputs an adaptively weighted summation of the inputs. Briefly, given $N$ input vectors $h_1,h_2,\cdots,h_N$, an attention layer outputs another $N$ vectors $\tilde{h}_1,\tilde{h}_2,\cdots,\tilde{h}_N$, where $\tilde{h}_i=\sum_{j=1}^{N}\alpha_jh_j$, and $\alpha_j$ is determined by the network. \texttt{FFN} layer is a network with one non-linear activation layer, which further processes $\tilde{h}_i$ and outputs an updated representation. We use $\Theta$ to denote all the parameters of the machine learning model.

We use a random variable $z$ to control the diverse generation for the same input. Given a protein binding pocket $\bx$, VAE generates a compound $\by$ according to the distribution $p(\mathbf{y}\vert \mathbf{x},z;\Theta)$. During the inference phase, $z$ is sampled from the multivariant standard Gaussian distribution $\mathcal{N}(0,I)$, and each $z$ leads to an output. During training, following the VAE criteria \citep{Bayes-VAE, CVAE, Generative-VAE}, $z$ is sampled from the VAE encoder $q(z\vert \mathbf{x},\mathbf{y})$, which is jointly learned with the generative model $\Theta$. 

The difference between our pocket encoder and the standard Transformer encoder is the way to obtain attention weights $\alpha_j$'s (see Figure \ref{fig:networl_arch}(b)). In our encoder, we use the product of the representation similarity (i.e., the similarity of $h_i$'s from the previous layer) and negative geometric distances (more specifically, $\exp(-{\rm distances}^2/\tau)$ where $\tau$ is a hyperparameter) to model the $\alpha$'s, whereas in standard Transformer, the distance is not modeled. For the decoder, the network architecture is the same as that in standard Transformer \cite{Transformer}. Considering the limited number of protein-ligand pairs, we use a pre-trained decoder: We randomly select $10M$ molecules from PubChem to train a SMILES-based language model (the objective function is to predict future tokens based on preceding tokens), and apply the pre-trained parameters to our decoder.   

\subsection{Results on DrugBank dataset}
We first evaluate the binding affinity and drug-likeness of the generated drugs. For this purpose, we built a test set based on the DrugBank~\citep{DrugBank}, which is a high-quality database with drugs and their target information. We obtained a dataset \ourDB{} with $1641$ target-ligand pairs and compared \ourM{} with two recently published baselines methods: 3DGen \cite{luo2021sbdd} and SECSE \cite{SystemicEvolutionary2021generating} (see Method Section for details). Due to the slow decoding speed of 3DGen and SECSE, we randomly sample $100$ target-ligand pairs from \ourDB{}. for each target, we generate compounds using 3DGen, SECSE and our method, and compare their performances. By default, the three methods all generate $20$ candidate compounds for each target.

\begin{figure}[!htbp]
\centering
\includegraphics[width=0.99\linewidth]{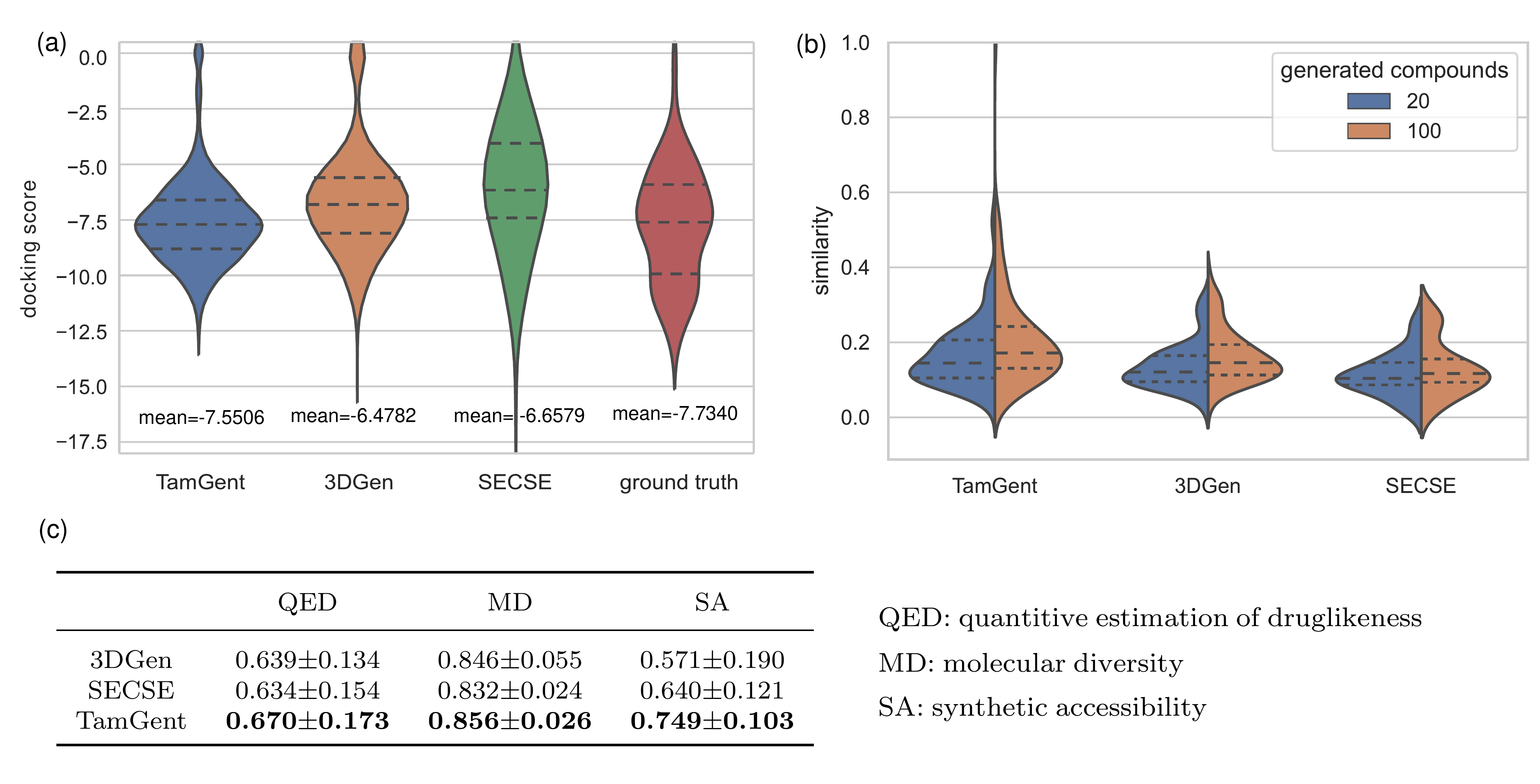}
\caption{\textbf{The properties of generated compounds.}  (a) The violin plot of  docking scores for compounds, as indicators for binding affinities. The three dashed lines in each part are the $25\%$, $50\%$ and $75\%$ quantiles. Compounds generated using our method (blue color) show stronger binding than those from other methods, while some drugs in the DrugBank have stronger binding interactions than all generated compounds (red color). (b) Similarity of the generated compounds to the ground truth drugs. The blue and yellow parts denote the similarity obtained by generating $20$ and $100$ compounds for each target. (c) The QED scores, molecular diversity scores and synthetic accessibility scores for compounds generated with different methods.}
\label{fig:drugbank_overall_results}
\end{figure}

For binding affinity evaluation, we use Autodock-smina\footnote{\url{https://sourceforge.net/projects/smina/}} to dock the generated molecules into binding sites and report the docking scores. The distributions of docking scores are shown in Figure~\ref{fig:drugbank_overall_results}(a). Compared with 3DGen and SECSE, the distribution of docking scores for compounds generated using \ourM{} shows an obvious downward shift, indicating stronger binding affinity.
 
It is noteworthy to point out that the distribution of ground truth set (i.e., the compounds from the Drugbank) is broader than generated molecules by all three methods, covering compounds with stronger binding to target proteins than generated compounds. This suggests that generative models can yield reasonable compounds that bind to the target, while there is still room for improvement to obtain stronger binding compounds.

We also analyze the relation between reference drugs and the generated compounds for each method. This is achieved by computing the similarity between the generate compounds and reference drugs in DrugBank defined by the Tanimoto similarity of Morgan fingerprints between a reference drug and a group of generated compounds. When each method is allowed to generate $20$ compounds, the average similarity achieved by \ourM{}, 3DGen and SECSE are $0.168$, $0.137$ and $0.113$ respectively. And when each method is allowed to generate $100$ compounds for each target, the average similarity metrics are $0.210$, $0.164$ and $0.131$. \ourM{} has the best performance in generating compounds similar to the drugs in the Drugbank. As shown in Figure \ref{fig:drugbank_overall_results}(b), the similarity distribution of \ourM{} shifts up compared with 3DGen and SECSE, in both cases with $20$ or $100$ generated compounds per target. On the other hand, the similarity is still at the low-side, indicating that the generated compounds are not necessarily the same as the known drugs.

To better evaluate model performance, following the common practice in machine learning research, we also calculate the following three metrics: (1) quatitative estimation of drug likeness (QED)~\citep{QED}; (2) molecular diversity (MD), which refers to the diversity between molecules generated for the same target; and (3) synthetic accessibility (SA) scores \cite{SAScore}, which estimates the synthetic feasibility of a compound. The results in Figure \ref{fig:drugbank_overall_results}(c) indicate that \ourM{} achieve the best performance.

After validating the effectiveness of our generative method, we decode the remaining $1541$ protein targets in \ourDB{}. 
We generate $100$ compounds for each target. After removing duplicate entries, \ourM{} yields $27040$ unique compounds in total, approximately $18$ compounds per target on average. Among them, $6233$ can be found in PubChem (downloaded on December 10, 2020, containing about $100$M compounds), and $20,807$ of them are new molecules that do not exist in PubChem, accounting for $76.9\%$ of the generated drugs. This shows that our method is capable of expanding compound libraries.

\begin{figure}[!htbp]
\centering
\includegraphics[width=0.8\linewidth]{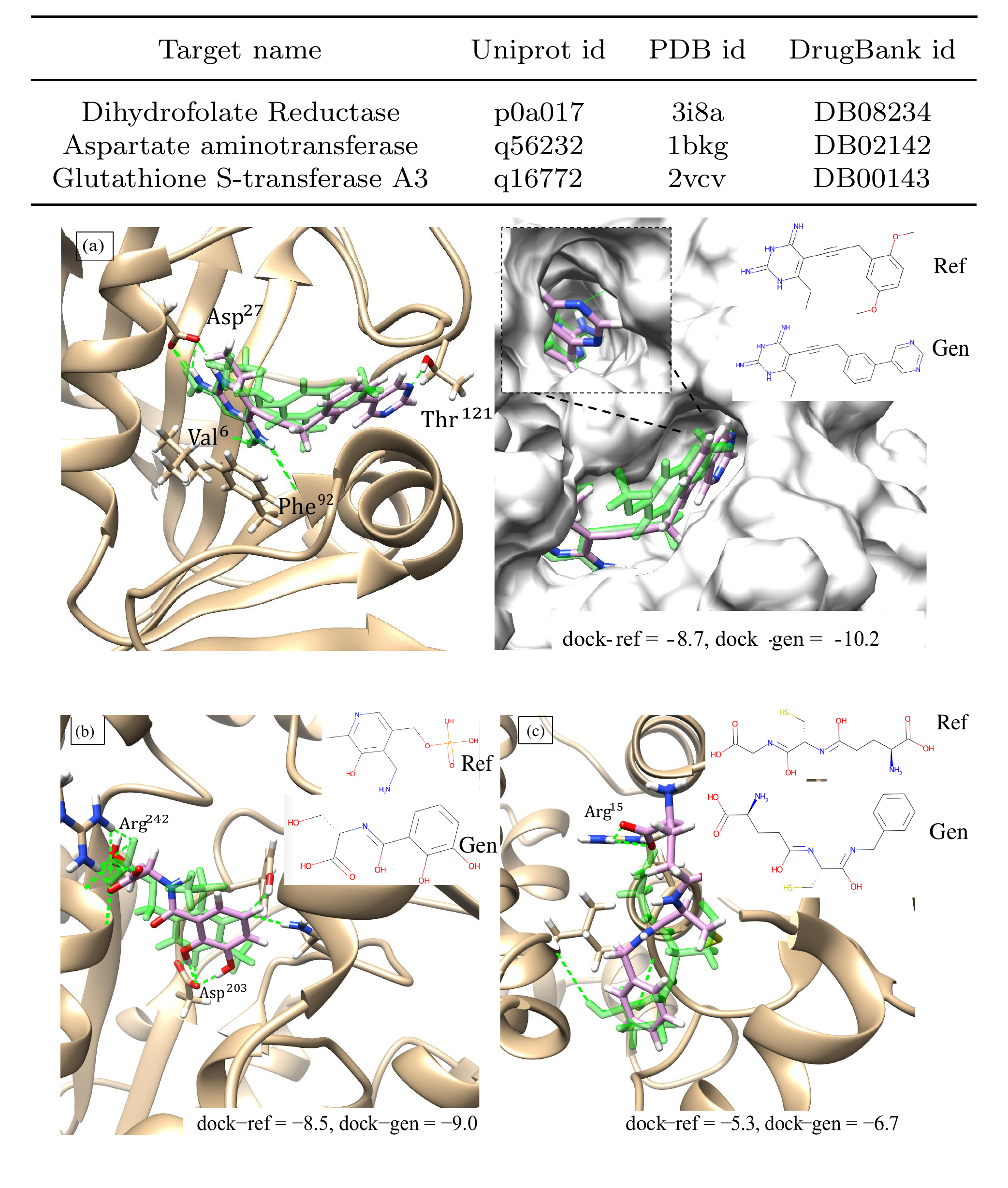}
\caption{\textbf{Representative cases for compound generations}. The target information is summarized in the table (top), where from left to right, the columns are target name, the uniprot id, and PDB id and the corresponding drug id in DrugBank. The docking poses are in the figure (bottom), where the proteins are shown by the ribbon by default. We also show the van der Waals surface of the first case for better visualizing pocket information.
In the docking pose, the reference drugs are shown in green color, and the generated molecules are colored according to atomic types. The 2D molecular graphs of the reference drug and generated compounds are shown in right panel of the corresponding subfigures. In these three cases, the reference drugs were found in the generated compounds. The predicted docking poses and scores suggest good quality of the de novo compounds generated using our method.}
\label{fig:case_DB_01}
\end{figure}

\ourM{} can generate both the reference drugs and de novo compounds. We observed the exact drug molecules among the generated compounds for $11$ drug targets extracted from the DrugBank. Their detailed information are in Supplementary Data A. Meanwhile, as aforementioned, the generated compounds can be  dissimilar to the known drugs. In order to assess the quality of such de novo drug compounds, we computed the docking scores for predicted docking poses and found some generated compounds showed better docking scores.  We further analyzed three representative target-compound pairs as shown in Figure \ref{fig:case_DB_01}.

For target Dihydrofolate Reductase, compared with the reference drug, the generated compound has an extra aromatic ring, and the nitrogen atom on this ring introduces an additional hydrogen bond with $\text{Thr}^{121}$, making it bound to the protein more tightly. When we use van der Waals surface to represent the protein, we can see that our new compound occupied more space of the pocket. Similarly, for target Aspartate aminotransferase and Glutathione S-transferase A3, we can generate new compounds with better docking scores.



\subsection{Case studies: SARS-CoV-2 main protease and KRAS G12C}
\subsubsection{SARS-CoV-2 main protease, \mpro{}}
Next, we focused on two targets that are under intense research: the SARS-CoV-2 main protease and the KRAS G12C mutant.
SARS-CoV-2 virus and the COVID-19 pandemic have severely affected everyone's life. Huge efforts have been put on to control this disease. SARS-CoV-2 main protease (briefly, \mpro) plays an important role in processing the polyproteins translated from viral RNA \cite{doi:10.1126/science.abb3405}, making a popular target for SARS-CoV-2 treatment. Thus, we collected $415$ high resolution structures by searching the keywords ``mpro sars-cov-2'' on PDBj\footnote{\url{https://pdbj.org/}}, and generated $4564$ unique compounds targeting \mpro{} with \ourM{}. The full list of the generated compounds can be found in Supplementary Data B.

We successfully recovered the GC-376 (DB15796) \cite{doi:10.1126/sciadv.abe0751}, a promising \mpro{} inhibitor candidate reported previously. We also generated six possible hit fragments for \mpro{} inhibitors verified by Douangamath et al. \cite{Douangamath2020-va} (PubChem (PDB) ids: 3695997 (5REK), 95851 (5RG0), 1353764 (5RFG), 19325375 (5REL), 3759658 (5REJ), 8405578 (5RFF)).


\begin{figure}[!htbp]
\centering
\subfigure[Visualization of generated drugs.]{
\includegraphics[width=0.8\linewidth]{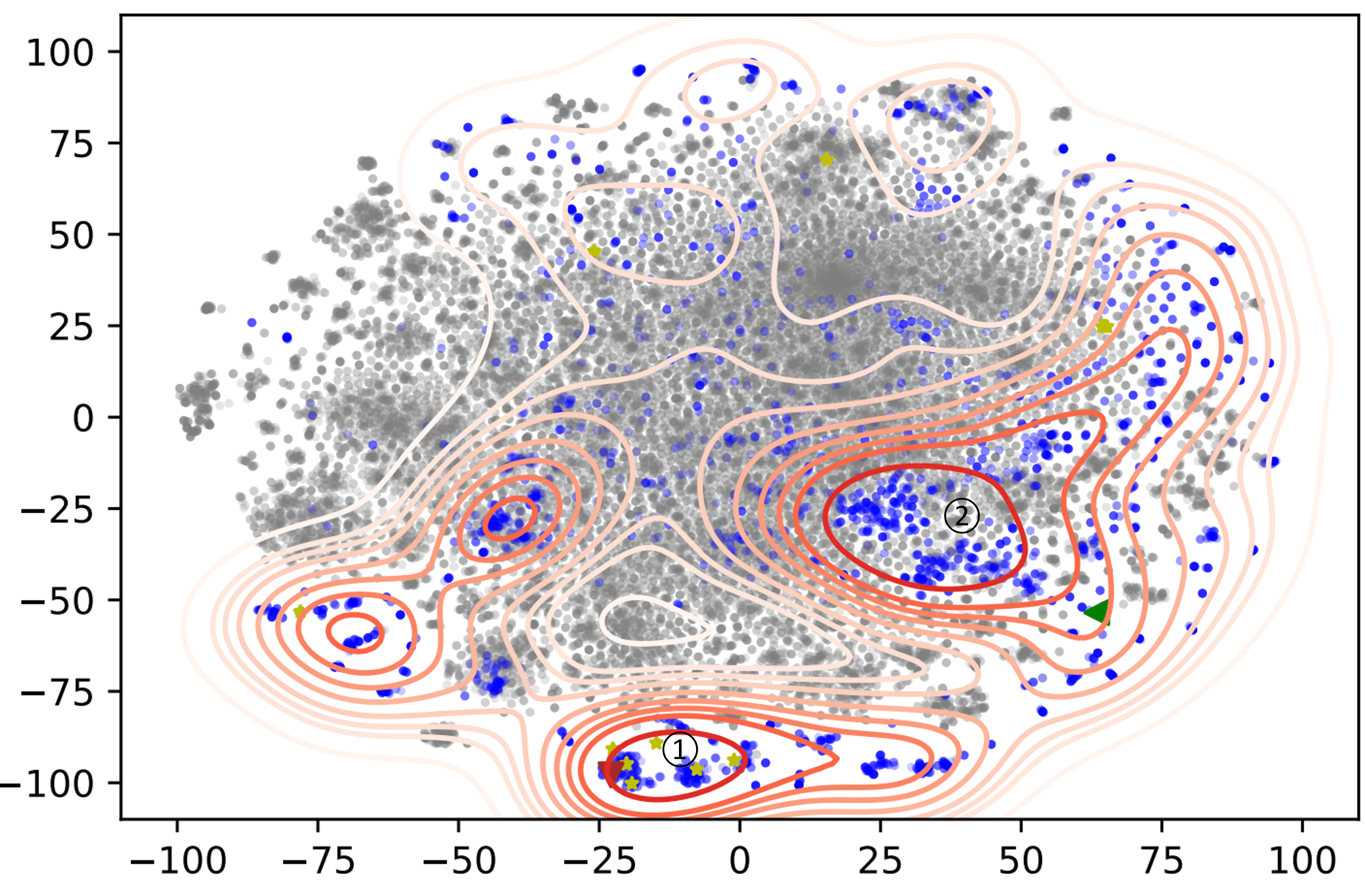}
}
\subfigure[Two examples of de novo candidates.]{
\includegraphics[width=0.8\linewidth]{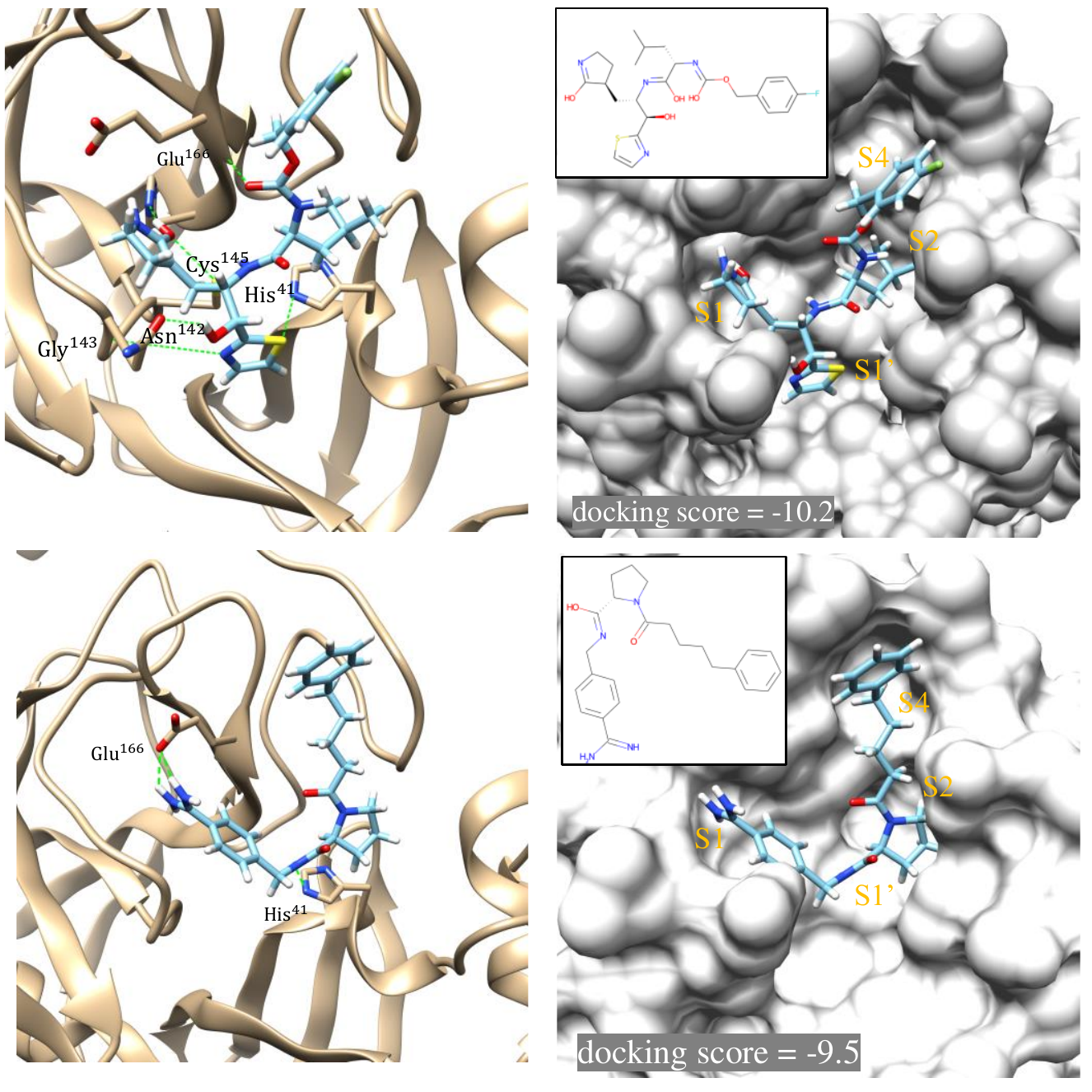}
}
\caption{\textbf{Analysis of our generated compounds.} (a) Visualization of all generated compounds for SARS-CoV-2 embedded in a 2D space. The gray circles are $30k$ randomly selected compounds from PubChem. Blue circles are our generated compounds based on \mpro{}, and the representative ones are highlighted using red and green triangles. Yellow stars denote possible \mpro{} inhibitors summarized by Mengist et al. \cite{mengist2021structural}. The contours are used to denote the density of our generated drugs (darker contours correspond to higher densities) (b) Two representative compounds generated using our method for \mpro{} target (top for the red triangle and bottom for the green triangle). The left panel denotes hydrogen-bonding interaction between generated compound and \mpro{}, and right panel denotes fitting to the pocket of the \mpro{}. 
The docking scores of the two compounds are $-10.2$ and $-9.5$ respectively.
}
\label{fig:analysis_generated_compounds}
\end{figure}

All $4564$ compounds generated by \ourM{} are embedded in a 2D space shown in Figure \ref{fig:analysis_generated_compounds}(a). Specifically, Morgan fingerprints were extracted for each compound and the dimensions were reduced to 2D using t-SNE algorithm. Blue dots are compounds generated using \ourM{}. Gray points are $30k$ compounds randomly selected from PubChem, which are supposed to have insignificant relevance to the \mpro{}. The contours indicate the density levels of the generated compounds for the \mpro{} in the space, with darker colors correspond to higher densities. The yellow stars are the possible \mpro{} inhibitors summarized by Mengist et al. \cite{mengist2021structural}, and the green and red triangles are the representative compounds generated by our method. 

We can observe that: (i) the randomly selected compounds and our generated compounds have different distributions. This shows that \ourM{} indeed generates compounds by considering the pocket information. (ii) A good number of generated compounds lie in one cluster (cluster $1$, Figure \ref{fig:analysis_generated_compounds}(a)). They share a similar structure whose Tanimoto similarity to GC-376 is $0.82$. More interestingly, six previously reported \mpro{} inhibitor candidates \cite{mengist2021structural} all lie in the same cluster. These results indicate the consistency between our generated compounds and reference inhibitors, as well as the capability of novel compound generation by \ourM{}. (iii) Similarly, we find a second cluster (cluster $2$, Figure \ref{fig:analysis_generated_compounds}(a)) with shared structure. Neither reference drugs nor compounds with promising docking scores were found within the cluster. We reason that the relatively small molecules in cluster $2$ may not be able to fill the $5$ sites at the binding pockets, since the average heavy atom numbers in cluster $1$ and cluster $2$ are $36$ and $16$ respectively. However, we still find some promising fragments in cluster 2 and leave the results in Supplementary Data C.

Next, we evaluated two diverse compounds generated by \ourM{} closely (Figure \ref{fig:analysis_generated_compounds}(b)). These compounds were estimated to have docking scores of $-10.2$ and $-9.5$ with \mpro{} respectively, outperforming known potential \mpro{} inhibitors (e.g., GC-376 with docking score $-9.4$). 
More importantly, these two generated compounds both fill the binding pocket. The two compounds can both fill S1 and S2 subsites, forming hydrogen bonds to the Glu$^{166}$ of the S1 subsite, and connect to the catalytic residue His$^{41}$ through hydrogen bonding interaction. In addition, the thiazole of the first compound has a hydrogen bond to the other catalytic residue Cys$^{145}$ and the Gly$^{143}$ in S1 subsite. From the results above, we believe that \ourM{} has generated promising hit compounds targeting \mpro{} from scratch.


\subsubsection{KRAS G12C mutant}
Targeted therapy aims at specifically attacking tumorous cells with mutated proteins instead of normal cells, and thus has become a desired paradigm for cancer treatment. Therefore, we deployed \ourM{} to generate candidate compounds for targeted therapy.

As a proof-of-concept, we focused on KRAS, a molecular switch that regulates cell growth, proliferation, differentiation, etc. KRAS mutations, including the well known G12C, will lead to malignant proliferation of cells, further promote the occurrence and development of tumors \cite{KRAS_function}. Thus, we generate candidate drugs that target KRAS G12C mutant specifically instead of wild-type KRAS with \ourM{} by selecting compounds with higher affinities to the mutant.

We collected $52$ crystal structures of G12C mutant targets from PDB and generated $580$ unique drugs in targeting peptide regions close to mutation sites. For each generated compound, we calculate docking scores to the G12C mutant and the scores to the wild-type KRAS. We also docked two KRAS-G12C inhibitors, Sotorasib (DB15569, approved by FDA) and Adagrassib (DB15568, submitted to FDA for approval), to G12C mutant and wild-type targets as references.

\begin{figure}[!htbp]
\centering
\subfigure[The histograms of docking scores.]{
\includegraphics[width=0.8\linewidth]{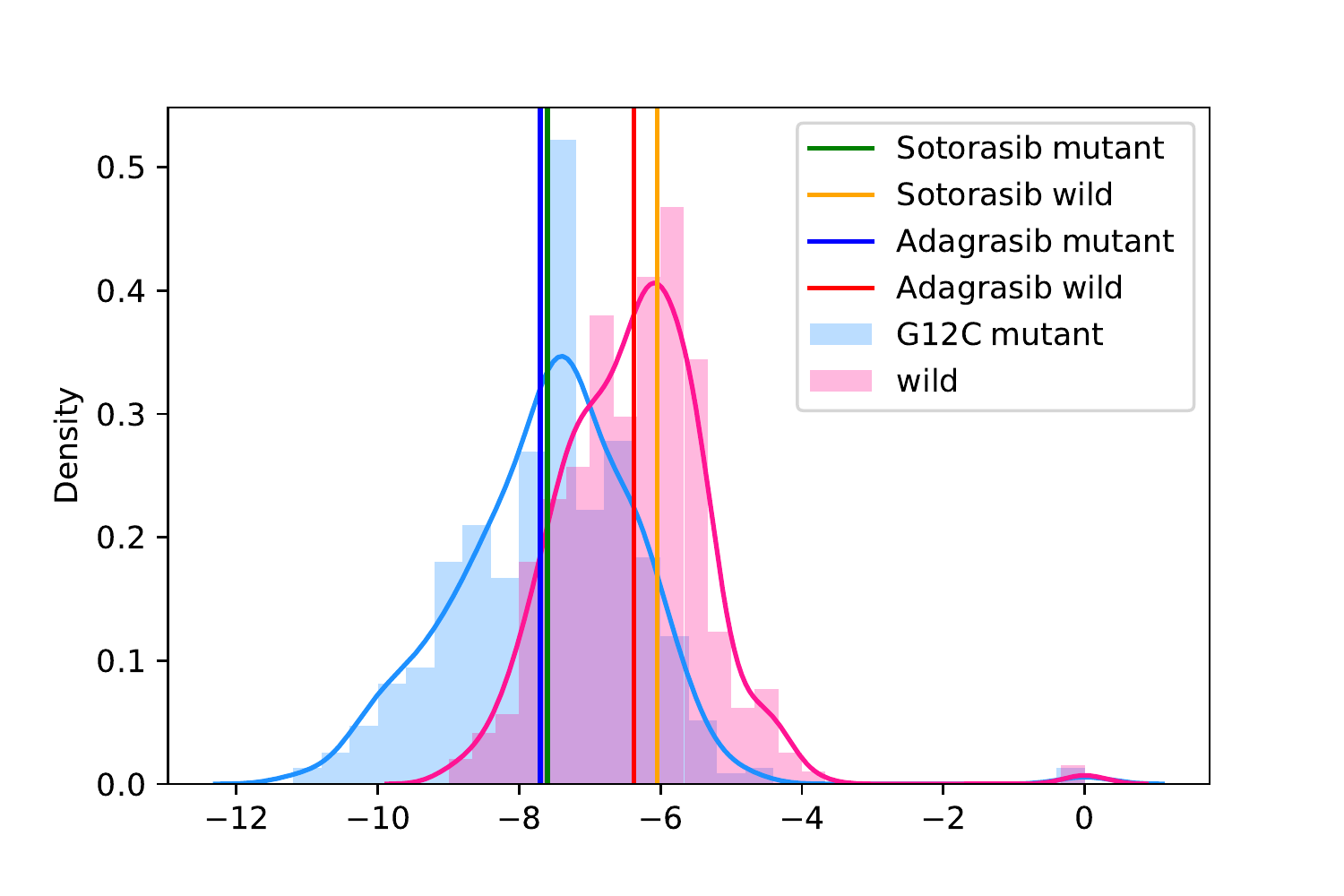}
}
\subfigure[Two examples of our generated compounds.]{\includegraphics[width=0.8\linewidth]{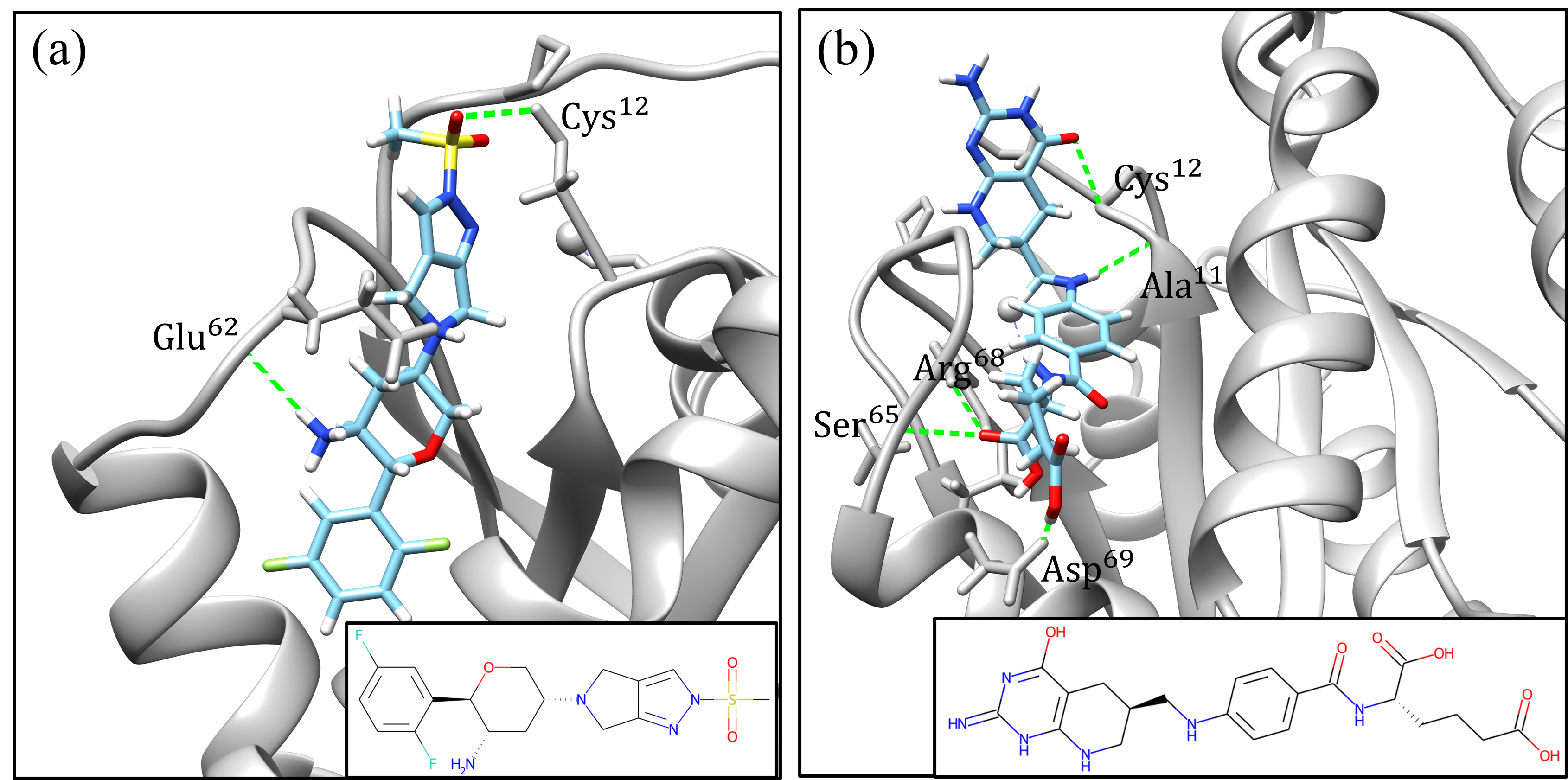}}
\caption{\textbf{Analysis of generated compounds for KRAS G12C.} (a) The blue and purple curves denote the docking scores with G12C mutant and the wile type proteins. The vertical lines are the docking scores of reference drugs as labelled in the legend. (b)  The docking poses with the mutate protein and the molecular graphs are visualized. The two compounds have hydrogen bonds with Cys$^{12}$ in the mutate protein, while do not have hydrogen bonds with the Gln$^{12}$ of the wild type. Let $D_m$ and $D_w$ denote the docking scores w.r.t. the mutate protein and wild type protein. For (a), $D_m=-7.3$, $D_w=-5.8$; for (b) $D_m=-8.7$, $D_w=-6.2$.}
\label{fig:KRAS_G12C_overall}
\end{figure}

Figure \ref{fig:KRAS_G12C_overall}(a) shows the distribution of docking scores. The blue and pink histograms/curves represent the docking results with G12C mutation target and wild-type target respectively. Clearly, the binding affinity of our generated drugs with the mutant target is better than that of the wild-type target. Docking scores of Sotorasib and Adagrassib are shown as vertical lines. $43.5\%$ of our generated drugs have better docking scores on mutant targets than Sotorasib and $40.2\%$ of them better than Adagrassib. 

We show two generated drugs in Figure \ref{fig:KRAS_G12C_overall}(b). Both compounds have better docking scores with the mutant than the wild type. In particular, they form hydrogen bonds to the cysteine (Cys$^{12}$) of the G12C mutant. In contrast, these crucial hydrogen bonds are not observed between the compounds and Gly$^{12}$ of the wild type KRAS. Unfortunately, we did not perform covalent docking due to the lack of domain knowledge, and the generated compounds are relatively small compared with Sotorasib and Adagrassib. However, we are exploring an improved model based on \ourM{} that aims at targeted drug generation with more domain knowledge taken into consideration.

\subsection{Computational efficiency}
Lastly, we evaluated the computational efficiency of \ourM{}. Our method is significantly faster than both 3DGen and SECSE when generating compound molecules for given targets, based on our experiments on a single Tesla P40 GPU. On average, to generate $20$ compounds for each target, 3DGen \cite{luo2021sbdd}, SECSE \cite{SystemicEvolutionary2021generating} take $25.4$ minutes and $30$ minutes respectively. In sharp contrast, \ourM{} takes only $0.3$ seconds to generate $20$ compounds with similar or better quality. These results show that our method is at least $5000$x faster than the above methods for the same task, demonstrating \ourM{}'s potential on high-throughput large-scale hit compounds generation for different targets. 


\section{Methods}
\subsection{Details of method and implementation}
Here, we describe the details about how to process the 3D input, the architectures of the encoder, decoder and the training objective functions. 

{\em Processing 3D input}: The amino acid $a_i$ $\forall i\in[N]$ is mapped to $d$-dimension vectors via an embedding layer $E_a$. The coordinate $r_i (i\in[N])$ is mapped to a $d$-dimension vector via a linear mapping. Considering we can rotate and translate a binding pocket while its spatial semantic information should preserve, we apply data augmentation to the coordinates. That is, in the input layer, for any $i\in[N]$,
\begin{equation}
h^{(0)}_i=E_aa_i + E_r\rho\left(r_i-\frac{1}{N}\sum_{j=1}^{N}r_j\right),
\end{equation}
where (i) $E_a$ and $E_r$ are learnable matrice, and they are optimized during model training; (ii) $\rho$ denotes a random roto-translation operation, and before using $\rho$, we center the coordinates to the origin. Thus we process the discrete input $\bx$ into $N$ continuous hidden representations $h^{(0)}_i$. 

{\em Pocket encoder}: The encoder stacks $L$ identical blocks. The output of the $l$-th block, i.e., $h^{(l)}_i$, is fed into the $(l+1)$-th layer for further processing and obtain $h^{(l+1)}_i$ for any $i\in[N]$ and $l\in\{0\}\cup[L-1]$. Each block consists of an attention layer and an \texttt{FFN} layer, which is a two-layer feed-forward network as that in the original Transformer \cite{Transformer}. To model the spatial distances of amino acids, we propose a new type of distance-aware attention. Mathematically,
\begin{equation}
\begin{aligned}
& \tilde{h}^{(l+1)}_i=\sum_{j=1}^N\alpha_j(W_vh^{(l)}_j),\\
&\alpha_j=\frac{\exp{\hat{\alpha}_j}}{\sum_{k=1}^{N}\exp{\hat{\alpha}_k}},\\
&\hat{\alpha}_j = \exp\left({-\frac{\Vert r_i-r_j \Vert ^2}{\tau}} \right) (h^{(l)\top}_iWh^{(l)}_j),
\end{aligned}
\end{equation}
where $W$ and $W_v$ are parameters to be optimized, and $\tau$ is the temperature hyperparameter to control. After that, $\tilde{h}^{(l+1)}_i$ is processed by an FFN layer and obtain 
\begin{equation}
h^{(l+1)}_i = \texttt{FFN}(\tilde{h}^{(l+1)}_i).
\end{equation}
The output from the last block, i.e., $h^{(L)}_{i}$ $\forall i\in[N]$, is the eventual representations of $\bx$ from the encoder. 

{\em The VAE encoder}: The VAE encoder models the posterior distribution of $z$ given a binding pocket $\bx$ and the corresponding ligand $\by$. The input of VAE encoder is defined as follows:
\begin{equation}
h^{(0)}_i=\left\{
	\begin{aligned}
	&E_aa_i + E_r\rho\left(r_i-\frac{1}{N}\sum_{j=1}^{N}r_j\right), \quad i\le N\\
	&E_yy_{i-N}, \quad i>N,
	\end{aligned}
\right.
\end{equation}
where $E_y$ is the embedding  of the SMILES. The VAE encoder follows the architecture of standard Transformer encoder \cite{Transformer}, which uses the vanilla self-attention layer rather than the distance-aware version due to the inavailability of the 3D ligand information.
The output from the last block, i.e., $h_i^{(L)}$ $\forall i \in [N]$, is mapped to the mean $\mu_{i}$ and covariance matrix $\Sigma_{i}$ of position $i$ via linear mapping, which can be used  for constructing $q(z\vert\mathbf{x}, \mathbf{y})$, by assuming $q(z\vert\bx,\by)$ is Gaussian. The ligand representations, i.e., $h^{(L)}_j$ $j>N$, are not used to construct $q(z\vert\bx,\by)$.



{\em Drug decoder}: The drug decoder is exactly the same as that in \cite{Transformer}, which consists of the self-attention layer, encoder-decoder attention layer (renamed as pocket-SMILES attention in Figure \ref{fig:networl_arch} and an FFN layer. The self-attention layer aggregates the representation from the previous block in the decoder, the pocket-SMILES attention processes the $h^{(L)}_i$ from the pocket encoder, and the \texttt{FFN} is exactly the same as that in the encoder. We pre-train the decoder on $10M$ compounds randomly selected from PubChem (denoted as $\mathcal{D}_0$) using the following objective function:
\begin{equation}
\min -\sum_{y\in\mathcal{D}_0}\frac{1}{M_y}\sum_{i=1}^{M_y}\log P(y_i \vert y_{i-1},y_{i-2},\cdots,y_1),
\end{equation}
where $M_y$ is the length of $y$. 

The pocket-SMILES attention layer takes all $h^{(L)}_i$ as inputs. Under the VAE variant, during training, the inputs are $h^{(L)}_i +z^\prime_i$, where $z^\prime_i$ is sampled from the distribution $q(z\vert \bx,\by)$ introduced above. During inference, the inputs are $h^{(L)}_i +z_i$ where $z_i$ is randomly sampled from $N(0,I)$. 

{\em Training}: The training objective is to minimize the following function:
\begin{equation}
\min_{\Theta,q}\frac{1}{\vert\mathcal{D} \vert}\sum_{(\bx,\by)\in\mathcal{D}}-\log P(\by\vert\bx,z;\Theta) + \beta\mathcal{D}_{\rm kl}\left(q(z\vert\bx,\by)\Vert p(z)\right),
\label{eq:vae_framework}
\end{equation}
where the $z$ in $\log P(\cdots)$ is sampled from $q(z\vert\bx,\by)$, $p(z)$ denotes the standard Gaussian distribution, and $\mathcal{D}_{\rm kl}$ denotes the KL divergence.

We use the data provided by Luo et al. \cite{luo2021sbdd} to train the models. The data is filtered from CrossDocked \cite{CrossDocked} and there are $123k$ target-ligand pairs (denoted as $\mathcal{D}$). 
We train our model with one NVIDIA V100 GPU. Both the pocket encoder and VAE encoder have $4$ layers with hidden dimension $256$.
The decoder has $12$ layers with hidden dimension $768$. We use Adam optimizer \cite{Adam} with initial learning $3\times{10}^{-5}$. 

\noindent{\bf Baselines}:  We mainly compare our method with two baselines methods 3DGen \cite{luo2021sbdd} and SECSE \cite{SystemicEvolutionary2021generating}.
\begin{enumerate}
\item 3DGen \cite{luo2021sbdd}, a representative deep learning baseline that uses a graph neural network to encode the 3D pocket information and direct generates the 3D conformation of candidate drugs. 
\item SECSE \cite{SystemicEvolutionary2021generating}, a hybrid system of de novo drug design with four steps: (1) build a fragment library; (2) generate compounds based on specific rules;
(3) filter unpromising molecules based on structures and properties; (4) evaluate the fitness of the current compounds based on docking scores or a deep-learning based model, by which we could select the compounds to be further optimized in the next round.
\end{enumerate}

\subsection{Procedure descriptions}

\noindent{\bf Construct the dataset}: There are three steps to build our dataset \ourDB{} based on DrugBank.
\begin{enumerate}
\item We download DrugBank dataset from its official website\footnote{\url{https://go.drugbank.com}} and parse it into target-drug pairs. We only keep the small-molecule drugs and filter out the pairs whose ligand is a protein. The targets and the drugs are represented by UniProt~\citep{UniProt} ids and SMILES strings respectively.
\item We query each collected UniProt id on the official
website~\footnote{\url{https://www.uniprot.org/uploadlists/}} and retrieve the corresponding PDB ids, by which we can get their 3D structures of the protein-ligand complex on the PDB \citep{PDB2021} website. We further split the complex into proteins and ligands. The $\by$ is the SMILES of the ligand and the $\bx$ is all amino acids within 10 angstrom from the ligand as input binding site. 
\item When the UniProt id and the splitted ligands of the protein-ligand pairs extracted from PDB are the same as those in DrugBank, we add them into a set \ourDB{}.
\end{enumerate}
We eventually obtain \ourDB{} with $1641$ target-ligand pairs after removing the redundant entries. Each target has $2.4$ PDB files on average.

\noindent{\bf Definition of similarity}:  Denote a reference target-ligand pair as $(t,d)$. Denote the generated compounds as $\mathcal{H}=\{\hat{h}_1,\hat{h}_2,\cdots,\hat{h}_n\}$. The similarity between $d$ and $\mathcal{H}$ is defined as $\max_{\hat{h}\in\mathcal{H}}\texttt{sim}(d,\hat{h})$, where $\texttt{sim}$ is implemented as Tanimoto similarity of Morgan fingerprints.

\noindent{\bf SARS-CoV-2 drug generation}: We use three models to generate candidate drugs for SARS-CoV-2 \mpro{}. The first model is the default generation model. The second model is finetuned on a subset of $300$ \mpro{} PDB files (denoted as $T_{\rm mpro}$). The third model is finetuned on a merged dataset of CrossDocked and  $T_{\rm mpro}$. Note that for the second and third models, we exclude the generated SMILES whose similarity is greater than $0.85$ to the SMILES in $T_{\rm mpro}$. 

\section{Discussion and conclusions}\label{sec:conclusion}
In this work, we propose a new model named \ourM{} for structure-based drug design that can directly generate candidate drugs based on target proteins. Our method can generate compounds that are similar to the known drugs, as well as novel compounds with potentially better binding affinities according to docking scores. Compared to previous methods, \ourM{} has better computational efficiency and compound quality.

The shortage of high quality target-drug interaction data imposes challenges to machine learning based methods. In this work, we leveraged the transformer-based approach and applied a pre-training model to obtain a representation of compound molecules. Using this approach, the generative model can decode compounds for the binding pocket of any given target. The diversity and drug likeliness of generative compounds are found to be superior than existing methods. 


The case studies on targets in the DrugBank reveal that the generated compounds can yield better binding scores than the reference drugs. Detailed examinations of docked complex structures provide physicochemical explanations for the enhanced binding interactions. Furthermore, the present method was tested on two widely studied cases: SARS-CoV-2 \mpro{} and KRAS G12C. The generated compounds have a good coverage of known inhibitors of \mpro{}, while generating new hypothetical compounds that potentially bind to the enzymatic sites. In the case of KRAS G12C, the generated compounds demonstrate stronger interactions with G12C mutant, compared to the wild type KRAS. This difference suggests that the compounds can interact with the mutated protein more specifically, while keeping normal cells without mutations less affected. In summary, the testing results demonstrate the potential of the present generative method. 

Despite the state-of-the-art performance, \ourM{} can be further improved from several aspects. Firstly, we are incorporating high-quality in-house target-drug pair data and patent data to boost our model. Secondly, we are involving pharmacological domain knowledge to refine generated compounds. For example, we expect our model to generate compounds with more stable or complex structure, as well as better ADME/T properties with feedback from domain experts. Lastly, we plan to focus on a few targets to fine-tune our model. For example, preliminary study indicates that we can significantly improve the candidate compounds by incorporating prior knowledge of \mpro{} or KRAS. In spite of current limitations, we believe that \ourM{} demonstrates an efficient solution to drug generation and that with future work \ourM{} will serve as an efficient generative framework to inspire novel drug discovery.


\backmatter

\bmhead{Supplementary information}
Supplementary Data A contains the generated compounds where the targets are from DrugBank. Supplementary Data B is the generated compounds for SARS-CoV-2 \mpro{}. Supplementary Data C is further analysis for SARS-CoV-2 \mpro{}.

\bmhead{Acknowledgments}
This work is conducted at Microsoft Research. The authors would like to thank Prof. Lifeng Fu for the helpful discussion on SARS-CoV-2.

\section*{Declarations}
\begin{itemize}
\item Funding: This work is supported by Microsoft Research.
\item Conflict of interest: None
\item Ethics approval: Not applicable.
\item Availability of data and materials: \url{https://github.com/HankerWu/TamGent}
\item Code availability: \url{https://github.com/HankerWu/TamGent}
\end{itemize}.

\begin{appendices}
\section{Evaluation metrics}
We evaluate the drug-likeness by the other metrics, including qualitative estimation of drug likeness (QED)~\citep{QED}, molecular diversity (MD) and synthetic accessibility (SA) \citep{SAScore}, which are widely reported in the machine learning literature. QED is calculated by \texttt{rdkit.Chem.QED}. The script to calculate SA score is in our github repository. Molecular diversity refers to the diversity between molecules generated for the same target, which is obtained by calculating the distance between the generated molecules, that is, $\text{MD}=\frac{1}{N( N-1)}\sum_{i\ne j}{1 - \mathrm{sim}(y_i, y_j)}$, where $N$ is the number of molecules generated, $\mathrm{sim}(y_i, y_j) $ is the Tanimoto similarity calculated from Morgan molecular fingerprints.

\end{appendices}




\begin{thebibliography}{31}
	\ifx \bisbn   \undefined \def \bisbn  #1{ISBN #1}\fi
	\ifx \binits  \undefined \def \binits#1{#1}\fi
	\ifx \bauthor  \undefined \def \bauthor#1{#1}\fi
	\ifx \batitle  \undefined \def \batitle#1{#1}\fi
	\ifx \bjtitle  \undefined \def \bjtitle#1{#1}\fi
	\ifx \bvolume  \undefined \def \bvolume#1{\textbf{#1}}\fi
	\ifx \byear  \undefined \def \byear#1{#1}\fi
	\ifx \bissue  \undefined \def \bissue#1{#1}\fi
	\ifx \bfpage  \undefined \def \bfpage#1{#1}\fi
	\ifx \blpage  \undefined \def \blpage #1{#1}\fi
	\ifx \burl  \undefined \def \burl#1{\textsf{#1}}\fi
	\ifx \doiurl  \undefined \def \doiurl#1{\url{https://doi.org/#1}}\fi
	\ifx \betal  \undefined \def \betal{\textit{et al.}}\fi
	\ifx \binstitute  \undefined \def \binstitute#1{#1}\fi
	\ifx \binstitutionaled  \undefined \def \binstitutionaled#1{#1}\fi
	\ifx \bctitle  \undefined \def \bctitle#1{#1}\fi
	\ifx \beditor  \undefined \def \beditor#1{#1}\fi
	\ifx \bpublisher  \undefined \def \bpublisher#1{#1}\fi
	\ifx \bbtitle  \undefined \def \bbtitle#1{#1}\fi
	\ifx \bedition  \undefined \def \bedition#1{#1}\fi
	\ifx \bseriesno  \undefined \def \bseriesno#1{#1}\fi
	\ifx \blocation  \undefined \def \blocation#1{#1}\fi
	\ifx \bsertitle  \undefined \def \bsertitle#1{#1}\fi
	\ifx \bsnm \undefined \def \bsnm#1{#1}\fi
	\ifx \bsuffix \undefined \def \bsuffix#1{#1}\fi
	\ifx \bparticle \undefined \def \bparticle#1{#1}\fi
	\ifx \barticle \undefined \def \barticle#1{#1}\fi
	\bibcommenthead
	\ifx \bconfdate \undefined \def \bconfdate #1{#1}\fi
	\ifx \botherref \undefined \def \botherref #1{#1}\fi
	\ifx \url \undefined \def \url#1{\textsf{#1}}\fi
	\ifx \bchapter \undefined \def \bchapter#1{#1}\fi
	\ifx \bbook \undefined \def \bbook#1{#1}\fi
	\ifx \bcomment \undefined \def \bcomment#1{#1}\fi
	\ifx \oauthor \undefined \def \oauthor#1{#1}\fi
	\ifx \citeauthoryear \undefined \def \citeauthoryear#1{#1}\fi
	\ifx \endbibitem  \undefined \def \endbibitem {}\fi
	\ifx \bconflocation  \undefined \def \bconflocation#1{#1}\fi
	\ifx \arxivurl  \undefined \def \arxivurl#1{\textsf{#1}}\fi
	\csname PreBibitemsHook\endcsname
	
	\bibitem{ProcessSBDD2003}
	\begin{barticle}
		\bauthor{\bsnm{Anderson}, \binits{A.C.}}:
		\batitle{The process of structure-based drug design}.
		\bjtitle{Chemistry \& biology}
		\bvolume{10}(\bissue{9}),
		\bfpage{787}--\blpage{797}
		(\byear{2003})
	\end{barticle}
	\endbibitem
	
	\bibitem{heifetz2022artificial}
	\begin{bbook}
		\bauthor{\bsnm{Heifetz}, \binits{A.}}:
		\bbtitle{Artificial Intelligence in Drug Design}.
		\bpublisher{Springer}, \blocation{???}
		(\byear{2022})
	\end{bbook}
	\endbibitem
	
	\bibitem{Bajorath2002}
	\begin{barticle}
		\bauthor{\bsnm{Bajorath}, \binits{J.}}:
		\batitle{Integration of virtual and high-throughput screening}.
		\bjtitle{Nature Reviews Drug Discovery}
		\bvolume{1}(\bissue{11}),
		\bfpage{882}--\blpage{894}
		(\byear{2002}).
		\doiurl{10.1038/nrd941}
	\end{barticle}
	\endbibitem
	
	\bibitem{Schneider2010}
	\begin{barticle}
		\bauthor{\bsnm{Schneider}, \binits{G.}}:
		\batitle{Virtual screening: an endless staircase?}
		\bjtitle{Nature Reviews Drug Discovery}
		\bvolume{9}(\bissue{4}),
		\bfpage{273}--\blpage{276}
		(\byear{2010}).
		\doiurl{10.1038/nrd3139}
	\end{barticle}
	\endbibitem
	
	\bibitem{doi:10.1021/ja512751q}
	\begin{barticle}
		\bauthor{\bsnm{Wang}, \binits{L.}},
		\bauthor{\bsnm{Wu}, \binits{Y.}},
		\bauthor{\bsnm{Deng}, \binits{Y.}},
		\bauthor{\bsnm{Kim}, \binits{B.}},
		\bauthor{\bsnm{Pierce}, \binits{L.}},
		\bauthor{\bsnm{Krilov}, \binits{G.}},
		\bauthor{\bsnm{Lupyan}, \binits{D.}},
		\bauthor{\bsnm{Robinson}, \binits{S.}},
		\bauthor{\bsnm{Dahlgren}, \binits{M.K.}},
		\bauthor{\bsnm{Greenwood}, \binits{J.}},
		\bauthor{\bsnm{Romero}, \binits{D.L.}},
		\bauthor{\bsnm{Masse}, \binits{C.}},
		\bauthor{\bsnm{Knight}, \binits{J.L.}},
		\bauthor{\bsnm{Steinbrecher}, \binits{T.}},
		\bauthor{\bsnm{Beuming}, \binits{T.}},
		\bauthor{\bsnm{Damm}, \binits{W.}},
		\bauthor{\bsnm{Harder}, \binits{E.}},
		\bauthor{\bsnm{Sherman}, \binits{W.}},
		\bauthor{\bsnm{Brewer}, \binits{M.}},
		\bauthor{\bsnm{Wester}, \binits{R.}},
		\bauthor{\bsnm{Murcko}, \binits{M.}},
		\bauthor{\bsnm{Frye}, \binits{L.}},
		\bauthor{\bsnm{Farid}, \binits{R.}},
		\bauthor{\bsnm{Lin}, \binits{T.}},
		\bauthor{\bsnm{Mobley}, \binits{D.L.}},
		\bauthor{\bsnm{Jorgensen}, \binits{W.L.}},
		\bauthor{\bsnm{Berne}, \binits{B.J.}},
		\bauthor{\bsnm{Friesner}, \binits{R.A.}},
		\bauthor{\bsnm{Abel}, \binits{R.}}:
		\batitle{Accurate and reliable prediction of relative ligand binding potency in
			prospective drug discovery by way of a modern free-energy calculation
			protocol and force field}.
		\bjtitle{Journal of the American Chemical Society}
		\bvolume{137}(\bissue{7}),
		\bfpage{2695}--\blpage{2703}
		(\byear{2015})
		{\href{https://arxiv.org/abs/https://doi.org/10.1021/ja512751q}{{https://doi.org/10.1021/ja512751q}}}.
		\doiurl{10.1021/ja512751q}
	\end{barticle}
	\endbibitem
	
	\bibitem{doi:10.1021/acs.jmedchem.5b01684}
	\begin{barticle}
		\bauthor{\bsnm{De~Vivo}, \binits{M.}},
		\bauthor{\bsnm{Masetti}, \binits{M.}},
		\bauthor{\bsnm{Bottegoni}, \binits{G.}},
		\bauthor{\bsnm{Cavalli}, \binits{A.}}:
		\batitle{Role of molecular dynamics and related methods in drug discovery}.
		\bjtitle{Journal of Medicinal Chemistry}
		\bvolume{59}(\bissue{9}),
		\bfpage{4035}--\blpage{4061}
		(\byear{2016}).
		\doiurl{10.1021/acs.jmedchem.5b01684}
	\end{barticle}
	\endbibitem
	
	\bibitem{Liu2018-ug}
	\begin{barticle}
		\bauthor{\bsnm{Liu}, \binits{X.}},
		\bauthor{\bsnm{Shi}, \binits{D.}},
		\bauthor{\bsnm{Zhou}, \binits{S.}},
		\bauthor{\bsnm{Liu}, \binits{H.}},
		\bauthor{\bsnm{Liu}, \binits{H.}},
		\bauthor{\bsnm{Yao}, \binits{X.}}:
		\batitle{Molecular dynamics simulations and novel drug discovery}.
		\bjtitle{Expert Opin. Drug Discov.}
		\bvolume{13}(\bissue{1}),
		\bfpage{23}--\blpage{37}
		(\byear{2018})
	\end{barticle}
	\endbibitem
	
	\bibitem{doi:10.1021/ar500432k}
	\begin{barticle}
		\bauthor{\bsnm{Reymond}, \binits{J.-L.}}:
		\batitle{The chemical space project}.
		\bjtitle{Accounts of Chemical Research}
		\bvolume{48}(\bissue{3}),
		\bfpage{722}--\blpage{730}
		(\byear{2015}).
		\doiurl{10.1021/ar500432k}
	\end{barticle}
	\endbibitem
	
	\bibitem{liGAN2020}
	\begin{botherref}
		\oauthor{\bsnm{Masuda}, \binits{T.}},
		\oauthor{\bsnm{Ragoza}, \binits{M.}},
		\oauthor{\bsnm{Koes}, \binits{D.R.}}:
		Generating 3d molecular structures conditional on a receptor binding site with
		deep generative models.
		arXiv preprint arXiv:2010.14442
		(2020)
	\end{botherref}
	\endbibitem
	
	\bibitem{StructureBasedMolGen2021}
	\begin{barticle}
		\bauthor{\bsnm{Ma}, \binits{B.}},
		\bauthor{\bsnm{Terayama}, \binits{K.}},
		\bauthor{\bsnm{Matsumoto}, \binits{S.}},
		\bauthor{\bsnm{Isaka}, \binits{Y.}},
		\bauthor{\bsnm{Sasakura}, \binits{Y.}},
		\bauthor{\bsnm{Iwata}, \binits{H.}},
		\bauthor{\bsnm{Araki}, \binits{M.}},
		\bauthor{\bsnm{Okuno}, \binits{Y.}}:
		\batitle{Structure-based de novo molecular generator combined with artificial
			intelligence and docking simulations}.
		\bjtitle{ChemRxiv}
		(\byear{2021}).
		\doiurl{10.26434/chemrxiv.14371967.v1}
	\end{barticle}
	\endbibitem
	
	\bibitem{luo2021sbdd}
	\begin{bchapter}
		\bauthor{\bsnm{Luo}, \binits{S.}},
		\bauthor{\bsnm{Guan}, \binits{J.}},
		\bauthor{\bsnm{Ma}, \binits{J.}},
		\bauthor{\bsnm{Peng}, \binits{J.}}:
		\bctitle{A 3d generative model for structure-based drug design}.
		In: \bbtitle{Advances in Neural Information Processing Systems},
		vol. \bseriesno{34}
		(\byear{2021})
	\end{bchapter}
	\endbibitem
	
	\bibitem{SystemicEvolutionary2021generating}
	\begin{barticle}
		\bauthor{\bsnm{Lu}, \binits{C.}},
		\bauthor{\bsnm{Liu}, \binits{S.}},
		\bauthor{\bsnm{Shi}, \binits{W.}},
		\bauthor{\bsnm{Yu}, \binits{J.}},
		\bauthor{\bsnm{Zhou}, \binits{Z.}},
		\bauthor{\bsnm{Zhang}, \binits{X.}},
		\bauthor{\bsnm{Lu}, \binits{X.}},
		\bauthor{\bsnm{Cai}, \binits{F.}},
		\bauthor{\bsnm{Xia}, \binits{N.}},
		\bauthor{\bsnm{Wang}, \binits{Y.}}:
		\batitle{Systemic evolutionary chemical space exploration for drug discovery}.
		\bjtitle{Journal of Cheminformatics}
		\bvolume{14}(\bissue{1}),
		\bfpage{19}
		(\byear{2022}).
		\doiurl{10.1186/s13321-022-00598-4}
	\end{barticle}
	\endbibitem
	
	\bibitem{CrossDocked}
	\begin{barticle}
		\bauthor{\bsnm{Francoeur}, \binits{P.G.}},
		\bauthor{\bsnm{Masuda}, \binits{T.}},
		\bauthor{\bsnm{Sunseri}, \binits{J.}},
		\bauthor{\bsnm{Jia}, \binits{A.}},
		\bauthor{\bsnm{Iovanisci}, \binits{R.B.}},
		\bauthor{\bsnm{Snyder}, \binits{I.}},
		\bauthor{\bsnm{Koes}, \binits{D.R.}}:
		\batitle{Three-dimensional convolutional neural networks and a cross-docked
			data set for structure-based drug design}.
		\bjtitle{Journal of Chemical Information and Modeling}
		\bvolume{60}(\bissue{9}),
		\bfpage{4200}--\blpage{4215}
		(\byear{2020})
	\end{barticle}
	\endbibitem
	
	\bibitem{EGCM}
	\begin{barticle}
		\bauthor{\bsnm{Xu}, \binits{M.}},
		\bauthor{\bsnm{Ran}, \binits{T.}},
		\bauthor{\bsnm{Chen}, \binits{H.}}:
		\batitle{De novo molecule design through the molecular generative model
			conditioned by 3d information of protein binding sites}.
		\bjtitle{Journal of Chemical Information and Modeling}
		\bvolume{61}(\bissue{7}),
		\bfpage{3240}--\blpage{3254}
		(\byear{2021})
	\end{barticle}
	\endbibitem
	
	\bibitem{Transformer}
	\begin{bchapter}
		\bauthor{\bsnm{Vaswani}, \binits{A.}},
		\bauthor{\bsnm{Shazeer}, \binits{N.}},
		\bauthor{\bsnm{Parmar}, \binits{N.}},
		\bauthor{\bsnm{Uszkoreit}, \binits{J.}},
		\bauthor{\bsnm{Jones}, \binits{L.}},
		\bauthor{\bsnm{Gomez}, \binits{A.N.}},
		\bauthor{\bsnm{Kaiser}, \binits{{\L}.}},
		\bauthor{\bsnm{Polosukhin}, \binits{I.}}:
		\bctitle{Attention is all you need}.
		In: \bbtitle{Advances in Neural Information Processing Systems},
		pp. \bfpage{5998}--\blpage{6008}
		(\byear{2017})
	\end{bchapter}
	\endbibitem
	
	\bibitem{Kim2021-th}
	\begin{barticle}
		\bauthor{\bsnm{Kim}, \binits{S.}},
		\bauthor{\bsnm{Chen}, \binits{J.}},
		\bauthor{\bsnm{Cheng}, \binits{T.}},
		\bauthor{\bsnm{Gindulyte}, \binits{A.}},
		\bauthor{\bsnm{He}, \binits{J.}},
		\bauthor{\bsnm{He}, \binits{S.}},
		\bauthor{\bsnm{Li}, \binits{Q.}},
		\bauthor{\bsnm{Shoemaker}, \binits{B.A.}},
		\bauthor{\bsnm{Thiessen}, \binits{P.A.}},
		\bauthor{\bsnm{Yu}, \binits{B.}},
		\bauthor{\bsnm{Zaslavsky}, \binits{L.}},
		\bauthor{\bsnm{Zhang}, \binits{J.}},
		\bauthor{\bsnm{Bolton}, \binits{E.E.}}:
		\batitle{{PubChem} in 2021: new data content and improved web interfaces}.
		\bjtitle{Nucleic Acids Res.}
		\bvolume{49}(\bissue{D1}),
		\bfpage{1388}--\blpage{1395}
		(\byear{2021})
	\end{barticle}
	\endbibitem
	
	\bibitem{Bayes-VAE}
	\begin{bchapter}
		\bauthor{\bsnm{Kingma}, \binits{D.P.}},
		\bauthor{\bsnm{Welling}, \binits{M.}}:
		\bctitle{Auto-encoding variational bayes}.
		In: \bbtitle{International Conference on Learning Representations}
		(\byear{2014})
	\end{bchapter}
	\endbibitem
	
	\bibitem{Koes2013-ta}
	\begin{barticle}
		\bauthor{\bsnm{Koes}, \binits{D.R.}},
		\bauthor{\bsnm{Baumgartner}, \binits{M.P.}},
		\bauthor{\bsnm{Camacho}, \binits{C.J.}}:
		\batitle{Lessons learned in empirical scoring with smina from the {CSAR} 2011
			benchmarking exercise}.
		\bjtitle{J. Chem. Inf. Model.}
		\bvolume{53}(\bissue{8}),
		\bfpage{1893}--\blpage{1904}
		(\byear{2013})
	\end{barticle}
	\endbibitem
	
	\bibitem{CVAE}
	\begin{botherref}
		\oauthor{\bsnm{Zhao}, \binits{T.}},
		\oauthor{\bsnm{Zhao}, \binits{R.}},
		\oauthor{\bsnm{Esk{\'{e}}nazi}, \binits{M.}}:
		Learning discourse-level diversity for neural dialog models using conditional
		variational autoencoders.
		CoRR
		\textbf{abs/1703.10960}
		(2017)
		{\href{https://arxiv.org/abs/1703.10960}{{1703.10960}}}
	\end{botherref}
	\endbibitem
	
	\bibitem{Generative-VAE}
	\begin{botherref}
		\oauthor{\bsnm{{Jimenez Rezende}}, \binits{D.}},
		\oauthor{\bsnm{{Mohamed}}, \binits{S.}},
		\oauthor{\bsnm{{Wierstra}}, \binits{D.}}:
		{Stochastic Backpropagation and Approximate Inference in Deep Generative
			Models}.
		arXiv e-prints,
		1401--4082
		(2014)
		{\href{https://arxiv.org/abs/1401.4082}{{arXiv:1401.4082}}}
		{[stat.ML]}
	\end{botherref}
	\endbibitem
	
	\bibitem{DrugBank}
	\begin{barticle}
		\bauthor{\bsnm{Wishart}, \binits{D.S.}},
		\bauthor{\bsnm{Feunang}, \binits{Y.D.}},
		\bauthor{\bsnm{Guo}, \binits{A.C.}},
		\bauthor{\bsnm{Lo}, \binits{E.J.}},
		\bauthor{\bsnm{Marcu}, \binits{A.}},
		\bauthor{\bsnm{Grant}, \binits{J.R.}},
		\bauthor{\bsnm{Sajed}, \binits{T.}},
		\bauthor{\bsnm{Johnson}, \binits{D.}},
		\bauthor{\bsnm{Li}, \binits{C.}},
		\bauthor{\bsnm{Sayeeda}, \binits{Z.}},
		\bauthor{\bsnm{Assempour}, \binits{N.}},
		\bauthor{\bsnm{Iynkkaran}, \binits{I.}},
		\bauthor{\bsnm{Liu}, \binits{Y.}},
		\bauthor{\bsnm{Maciejewski}, \binits{A.}},
		\bauthor{\bsnm{Gale}, \binits{N.}},
		\bauthor{\bsnm{Wilson}, \binits{A.}},
		\bauthor{\bsnm{Chin}, \binits{L.}},
		\bauthor{\bsnm{Cummings}, \binits{R.}},
		\bauthor{\bsnm{Le}, \binits{D.}},
		\bauthor{\bsnm{Pon}, \binits{A.}},
		\bauthor{\bsnm{Knox}, \binits{C.}},
		\bauthor{\bsnm{Wilson}, \binits{M.}}:
		\batitle{{DrugBank 5.0: a major update to the DrugBank database for 2018}}.
		\bjtitle{Nucleic Acids Research}
		\bvolume{46}(\bissue{D1}),
		\bfpage{1074}--\blpage{1082}
		(\byear{2017})
		{\href{https://arxiv.org/abs/https://academic.oup.com/nar/article-pdf/46/D1/D1074/23162116/gkx1037.pdf}{{https://academic.oup.com/nar/article-pdf/46/D1/D1074/23162116/gkx1037.pdf}}}.
		\doiurl{10.1093/nar/gkx1037}
	\end{barticle}
	\endbibitem
	
	\bibitem{QED}
	\begin{barticle}
		\bauthor{\bsnm{Bickerton}, \binits{G.R.}},
		\bauthor{\bsnm{Paolini}, \binits{G.V.}},
		\bauthor{\bsnm{Besnard}, \binits{J.}},
		\bauthor{\bsnm{Muresan}, \binits{S.}},
		\bauthor{\bsnm{Hopkins}, \binits{A.L.}}:
		\batitle{Quantifying the chemical beauty of drugs}.
		\bjtitle{Nature chemistry}
		\bvolume{4}(\bissue{2}),
		\bfpage{90}--\blpage{98}
		(\byear{2012})
	\end{barticle}
	\endbibitem
	
	\bibitem{SAScore}
	\begin{barticle}
		\bauthor{\bsnm{Ertl}, \binits{P.}},
		\bauthor{\bsnm{Schuffenhauer}, \binits{A.}}:
		\batitle{Estimation of synthetic accessibility score of drug-like molecules
			based on molecular complexity and fragment contributions}.
		\bjtitle{Journal of cheminformatics}
		\bvolume{1}(\bissue{1}),
		\bfpage{1}--\blpage{11}
		(\byear{2009})
	\end{barticle}
	\endbibitem
	
	\bibitem{doi:10.1126/science.abb3405}
	\begin{barticle}
		\bauthor{\bsnm{Zhang}, \binits{L.}},
		\bauthor{\bsnm{Lin}, \binits{D.}},
		\bauthor{\bsnm{Sun}, \binits{X.}},
		\bauthor{\bsnm{Curth}, \binits{U.}},
		\bauthor{\bsnm{Drosten}, \binits{C.}},
		\bauthor{\bsnm{Sauerhering}, \binits{L.}},
		\bauthor{\bsnm{Becker}, \binits{S.}},
		\bauthor{\bsnm{Rox}, \binits{K.}},
		\bauthor{\bsnm{Hilgenfeld}, \binits{R.}}:
		\batitle{Crystal structure of sars-cov-2 main protease provides a basis for
			design of improved \&\#x3b1;-ketoamide inhibitors}.
		\bjtitle{Science}
		\bvolume{368}(\bissue{6489}),
		\bfpage{409}--\blpage{412}
		(\byear{2020}).
		\doiurl{10.1126/science.abb3405}
	\end{barticle}
	\endbibitem
	
	\bibitem{doi:10.1126/sciadv.abe0751}
	\begin{barticle}
		\bauthor{\bsnm{Sacco}, \binits{M.D.}},
		\bauthor{\bsnm{Ma}, \binits{C.}},
		\bauthor{\bsnm{Lagarias}, \binits{P.}},
		\bauthor{\bsnm{Gao}, \binits{A.}},
		\bauthor{\bsnm{Townsend}, \binits{J.A.}},
		\bauthor{\bsnm{Meng}, \binits{X.}},
		\bauthor{\bsnm{Dube}, \binits{P.}},
		\bauthor{\bsnm{Zhang}, \binits{X.}},
		\bauthor{\bsnm{Hu}, \binits{Y.}},
		\bauthor{\bsnm{Kitamura}, \binits{N.}},
		\bauthor{\bsnm{Hurst}, \binits{B.}},
		\bauthor{\bsnm{Tarbet}, \binits{B.}},
		\bauthor{\bsnm{Marty}, \binits{M.T.}},
		\bauthor{\bsnm{Kolocouris}, \binits{A.}},
		\bauthor{\bsnm{Xiang}, \binits{Y.}},
		\bauthor{\bsnm{Chen}, \binits{Y.}},
		\bauthor{\bsnm{Wang}, \binits{J.}}:
		\batitle{Structure and inhibition of the sars-cov-2 main protease reveal
			strategy for developing dual inhibitors against m<sup>pro</sup> and cathepsin
			l}.
		\bjtitle{Science Advances}
		\bvolume{6}(\bissue{50}),
		\bfpage{0751}
		(\byear{2020})
		{\href{https://arxiv.org/abs/https://www.science.org/doi/pdf/10.1126/sciadv.abe0751}{{https://www.science.org/doi/pdf/10.1126/sciadv.abe0751}}}.
		\doiurl{10.1126/sciadv.abe0751}
	\end{barticle}
	\endbibitem
	
	\bibitem{Douangamath2020-va}
	\begin{barticle}
		\bauthor{\bsnm{Douangamath}, \binits{A.}},
		\bauthor{\bsnm{Fearon}, \binits{D.}},
		\bauthor{\bsnm{Gehrtz}, \binits{P.}},
		\bauthor{\bsnm{Krojer}, \binits{T.}},
		\bauthor{\bsnm{Lukacik}, \binits{P.}},
		\bauthor{\bsnm{Owen}, \binits{C.D.}},
		\bauthor{\bsnm{Resnick}, \binits{E.}},
		\bauthor{\bsnm{Strain-Damerell}, \binits{C.}},
		\bauthor{\bsnm{Aimon}, \binits{A.}},
		\bauthor{\bsnm{{\'A}br{\'a}nyi-Balogh}, \binits{P.}},
		\bauthor{\bsnm{Brand{\~a}o-Neto}, \binits{J.}},
		\bauthor{\bsnm{Carbery}, \binits{A.}},
		\bauthor{\bsnm{Davison}, \binits{G.}},
		\bauthor{\bsnm{Dias}, \binits{A.}},
		\bauthor{\bsnm{Downes}, \binits{T.D.}},
		\bauthor{\bsnm{Dunnett}, \binits{L.}},
		\bauthor{\bsnm{Fairhead}, \binits{M.}},
		\bauthor{\bsnm{Firth}, \binits{J.D.}},
		\bauthor{\bsnm{Jones}, \binits{S.P.}},
		\bauthor{\bsnm{Keeley}, \binits{A.}},
		\bauthor{\bsnm{Keser{\"u}}, \binits{G.M.}},
		\bauthor{\bsnm{Klein}, \binits{H.F.}},
		\bauthor{\bsnm{Martin}, \binits{M.P.}},
		\bauthor{\bsnm{Noble}, \binits{M.E.M.}},
		\bauthor{\bsnm{O'Brien}, \binits{P.}},
		\bauthor{\bsnm{Powell}, \binits{A.}},
		\bauthor{\bsnm{Reddi}, \binits{R.N.}},
		\bauthor{\bsnm{Skyner}, \binits{R.}},
		\bauthor{\bsnm{Snee}, \binits{M.}},
		\bauthor{\bsnm{Waring}, \binits{M.J.}},
		\bauthor{\bsnm{Wild}, \binits{C.}},
		\bauthor{\bsnm{London}, \binits{N.}},
		\bauthor{\bparticle{von} \bsnm{Delft}, \binits{F.}},
		\bauthor{\bsnm{Walsh}, \binits{M.A.}}:
		\batitle{Crystallographic and electrophilic fragment screening of the
			{SARS-CoV-2} main protease}.
		\bjtitle{Nat. Commun.}
		\bvolume{11}(\bissue{1}),
		\bfpage{5047}
		(\byear{2020})
	\end{barticle}
	\endbibitem
	
	\bibitem{mengist2021structural}
	\begin{barticle}
		\bauthor{\bsnm{Mengist}, \binits{H.M.}},
		\bauthor{\bsnm{Dilnessa}, \binits{T.}},
		\bauthor{\bsnm{Jin}, \binits{T.}}:
		\batitle{Structural basis of potential inhibitors targeting sars-cov-2 main
			protease}.
		\bjtitle{Frontiers in Chemistry}
		\bvolume{9},
		\bfpage{622898}
		(\byear{2021})
	\end{barticle}
	\endbibitem
	
	\bibitem{KRAS_function}
	\begin{barticle}
		\bauthor{\bsnm{Yun}, \binits{J.}},
		\bauthor{\bsnm{Rago}, \binits{C.}},
		\bauthor{\bsnm{Cheong}, \binits{I.}},
		\bauthor{\bsnm{Pagliarini}, \binits{R.}},
		\bauthor{\bsnm{Angenendt}, \binits{P.}},
		\bauthor{\bsnm{Rajagopalan}, \binits{H.}},
		\bauthor{\bsnm{Schmidt}, \binits{K.}},
		\bauthor{\bsnm{Willson}, \binits{J.K.V.}},
		\bauthor{\bsnm{Markowitz}, \binits{S.}},
		\bauthor{\bsnm{Zhou}, \binits{S.}},
		\bauthor{\bsnm{Diaz}, \binits{L.A.}},
		\bauthor{\bsnm{Velculescu}, \binits{V.E.}},
		\bauthor{\bsnm{Lengauer}, \binits{C.}},
		\bauthor{\bsnm{Kinzler}, \binits{K.W.}},
		\bauthor{\bsnm{Vogelstein}, \binits{B.}},
		\bauthor{\bsnm{Papadopoulos}, \binits{N.}}:
		\batitle{Glucose deprivation contributes to the development of <i>kras</i>
			pathway mutations in tumor cells}.
		\bjtitle{Science}
		\bvolume{325}(\bissue{5947}),
		\bfpage{1555}--\blpage{1559}
		(\byear{2009})
		{\href{https://arxiv.org/abs/https://www.science.org/doi/pdf/10.1126/science.1174229}{{https://www.science.org/doi/pdf/10.1126/science.1174229}}}.
		\doiurl{10.1126/science.1174229}
	\end{barticle}
	\endbibitem
	
	\bibitem{Adam}
	\begin{bchapter}
		\bauthor{\bsnm{Kingma}, \binits{D.P.}},
		\bauthor{\bsnm{Ba}, \binits{J.}}:
		\bctitle{Adam: {A} method for stochastic optimization}.
		In: \beditor{\bsnm{Bengio}, \binits{Y.}},
		\beditor{\bsnm{LeCun}, \binits{Y.}} (eds.)
		\bbtitle{3rd International Conference on Learning Representations, {ICLR} 2015,
			San Diego, CA, USA, May 7-9, 2015, Conference Track Proceedings}
		(\byear{2015}).
		\burl{http://arxiv.org/abs/1412.6980}
	\end{bchapter}
	\endbibitem
	
	\bibitem{UniProt}
	\begin{barticle}
		\bauthor{\bsnm{Consortium}, \binits{T.U.}}:
		\batitle{{UniProt: the universal protein knowledgebase in 2021}}.
		\bjtitle{Nucleic Acids Research}
		\bvolume{49}(\bissue{D1}),
		\bfpage{480}--\blpage{489}
		(\byear{2020})
		{\href{https://arxiv.org/abs/https://academic.oup.com/nar/article-pdf/49/D1/D480/35364103/gkaa1100.pdf}{{https://academic.oup.com/nar/article-pdf/49/D1/D480/35364103/gkaa1100.pdf}}}.
		\doiurl{10.1093/nar/gkaa1100}
	\end{barticle}
	\endbibitem
	
	\bibitem{PDB2021}
	\begin{barticle}
		\bauthor{\bsnm{Burley}, \binits{S.K.}},
		\bauthor{\bsnm{Bhikadiya}, \binits{C.}},
		\bauthor{\bsnm{Bi}, \binits{C.}},
		\bauthor{\bsnm{Bittrich}, \binits{S.}},
		\bauthor{\bsnm{Chen}, \binits{L.}},
		\bauthor{\bsnm{Crichlow}, \binits{G.V.}},
		\bauthor{\bsnm{Christie}, \binits{C.H.}},
		\bauthor{\bsnm{Dalenberg}, \binits{K.}},
		\bauthor{\bsnm{Di~Costanzo}, \binits{L.}},
		\bauthor{\bsnm{Duarte}, \binits{J.M.}},
		\bauthor{\bsnm{Dutta}, \binits{S.}},
		\bauthor{\bsnm{Feng}, \binits{Z.}},
		\bauthor{\bsnm{Ganesan}, \binits{S.}},
		\bauthor{\bsnm{Goodsell}, \binits{D.S.}},
		\bauthor{\bsnm{Ghosh}, \binits{S.}},
		\bauthor{\bsnm{Green}, \binits{R.K.}},
		\bauthor{\bsnm{Guranović}, \binits{V.}},
		\bauthor{\bsnm{Guzenko}, \binits{D.}},
		\bauthor{\bsnm{Hudson}, \binits{B.P.}},
		\bauthor{\bsnm{Lawson}, \binits{C.L.}},
		\bauthor{\bsnm{Liang}, \binits{Y.}},
		\bauthor{\bsnm{Lowe}, \binits{R.}},
		\bauthor{\bsnm{Namkoong}, \binits{H.}},
		\bauthor{\bsnm{Peisach}, \binits{E.}},
		\bauthor{\bsnm{Persikova}, \binits{I.}},
		\bauthor{\bsnm{Randle}, \binits{C.}},
		\bauthor{\bsnm{Rose}, \binits{A.}},
		\bauthor{\bsnm{Rose}, \binits{Y.}},
		\bauthor{\bsnm{Sali}, \binits{A.}},
		\bauthor{\bsnm{Segura}, \binits{J.}},
		\bauthor{\bsnm{Sekharan}, \binits{M.}},
		\bauthor{\bsnm{Shao}, \binits{C.}},
		\bauthor{\bsnm{Tao}, \binits{Y.-P.}},
		\bauthor{\bsnm{Voigt}, \binits{M.}},
		\bauthor{\bsnm{Westbrook}, \binits{J.D.}},
		\bauthor{\bsnm{Young}, \binits{J.Y.}},
		\bauthor{\bsnm{Zardecki}, \binits{C.}},
		\bauthor{\bsnm{Zhuravleva}, \binits{M.}}:
		\batitle{{RCSB Protein Data Bank: powerful new tools for exploring 3D
				structures of biological macromolecules for basic and applied research and
				education in fundamental biology, biomedicine, biotechnology, bioengineering
				and energy sciences}}.
		\bjtitle{Nucleic Acids Research}
		\bvolume{49}(\bissue{D1}),
		\bfpage{437}--\blpage{451}
		(\byear{2020})
		{\href{https://arxiv.org/abs/https://academic.oup.com/nar/article-pdf/49/D1/D437/35364241/gkaa1038.pdf}{{https://academic.oup.com/nar/article-pdf/49/D1/D437/35364241/gkaa1038.pdf}}}.
		\doiurl{10.1093/nar/gkaa1038}
	\end{barticle}
	\endbibitem
	
\end{thebibliography}


\end{document}